# An information services algorithm to heuristically summarize IP addresses for a distributed, hierarchical directory service

Marcos Portnoi, Martin Swany
Department of Computer and Information Sciences
University of Delaware
Newark, DE 19716, U.S.A.
{portnoi, swany}@cis.udel.edu

Jason Zurawski
Internet2
Washington, DC 20036, U.S.A.
zurawski@internet2.edu

*Abstract*— A distributed, hierarchical information service for computer networks might use several service instances, located in different layers. A distributed directory service, for example, might be comprised of upper level listings, and local directories. The upper level listings contain a compact version of the local directories. Clients desiring to access the information contained in local directories might first access the high-level listings, in order to locate the appropriate local instance. One of the keys for the competent operation of such service is the ability of properly summarizing the information which will be maintained in the upper level directories. We analyze the case of the Lookup Service in the Information Services plane of perfSONAR performance monitoring distributed architecture, which implements IP address summarization in its functions. We propose an empirical method, or heuristic, to perform the summarizations, based on the PATRICIA tree. We further apply the heuristic on a simulated distributed test bed and examine the results.

## I. INTRODUCTION

Certain distributed computer network information services work by dispersing resources and data among several instances of the service, in a hierarchical manner. One such service can be a distributed, hierarchical directory service, where "lower-level" instances maintain local data and publish the data to "higher-level" instances. The higher-level instances are responsible for keeping a compact listing of all the data administrated by lower level instances, and must be able to indicate which particular instance maintains specific data. The higher-level instances, therefore, hold a summary of all the lower level counterparts.

For this mechanism to operate efficiently, the lower level instances must summarize their data and publish this summarization to the upper level layer. The kind of data being summarized might accommodate different techniques for this procedure, some of them based on heuristics. One such example is the Lookup Service in the Information Services plane used in perfSONAR architecture [1], which resembles a distributed directory with two levels. Its lower level instances summarize the controlled data and forward the compacted version to the upper layer. Among the published information, there are data descriptors (metadata) and IPv4 addresses.

This paper describes the work done in producing a heuristic to generate IPv4 addresses summarizations, and utilizes the realm of perfSONAR's Information Services plane as a motivator. The document is divided as follows. Section II delineates the perfSONAR architecture, the Information Services plane, and stimulates the need for IP summarization. Section III demonstrates the heuristic's mechanism, studying the construction of possible summaries and the decision steps. In Section IV, we employ the heuristic on several test sets and analyze the results. Section V brings considerations about other summarization techniques and route aggregation algorithms. Finally, in Section VI, we conclude the work and present our final thoughts.

## II. INFORMATION SERVICES IN PERFSONAR

Oriented to network performance monitoring, perfSONAR [1] is a distributed, services oriented architecture, formulated by a worldwide consortium of organizations, and it is comprised by a set of protocols and interoperable software packages [8][14]. The purpose of perfSONAR is to collect, store, and publish network-monitoring data, such as latency, topology, utilization, as also aid in diagnosing performance issues and anomalies.

The architecture defines several *service types*, namely Measurement Point (MP) service; Measurement Archive (MA) service; Transformation Service (TS); Lookup Service (LS); Topology Service (ToS); Authentication Service (AS); and Resource Protector Service (RPS) [1]. In particular, the LS serves as a registering hub [16] for all participating services and the capabilities they furnish. We will concentrate in this service for the purposes of motivating this paper.

The *protocols*, based on SOAP XML messages [7], regulate how the services communicate. The *software packages* act as middleware between the performance measurement tools and applications for diagnostic and visualization; they implement the service types and guarantee that they work across multiple or multi-domain networks.

### A. The Lookup Service

#### 1) General Operation
The general operation of the LS in perfSONAR context is depicted in Figure 1.

Essentially, the LS acts as a distributed directory. Currently, there are two types of instances (and levels) of LSs in commission in the network: the *global Lookup Service*, or gLS; and the *home Lookup Service*, or hLS. The instances are typically individual network devices, or computers, running the LS service in either gLS or hLS mode. The gLSs compose the upper LS level, and the hLSs form the lower level.

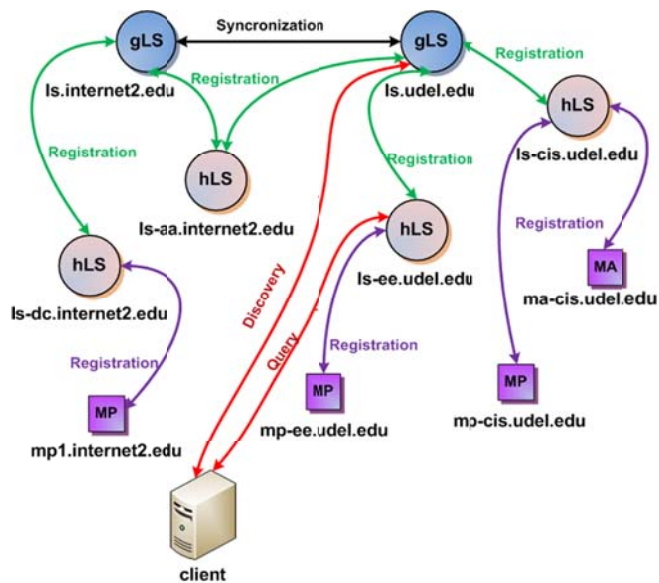

Figure 1. General operation of LS instances in perfSONAR's Information Services plane.

New hosts that wish to provide some monitoring service to the perfSONAR domain(s) (such as an MA or MP service) must register themselves to some hLS. These new services find a hLS by either knowing its URI (Uniform Resource Identifier) beforehand, or by performing a *discovery query* to the upper level gLS, which will in turn reply with existing hLS instances. Obviously, the new service must know where to find at least one gLS; there is a list of gLSs, the *gLS root hints* file, hosted by well-known servers that provide the URIs for available gLSs.

Each hLS keeps account of the registrations of the individual services that registered with this particular hLS. Each hLS must periodically register itself with at least one upper level gLS, so that the gLSs can update their information about current hLS instances. Finally, the gLSs synchronize with other gLSs. This layered, distributed mechanism attains scalability advantages [2]. We examine this concept next.

*2) LS: Keeping and Publishing Information (and Motivating Summarization)*

When a new service registers with a hLS, it conveys information such as the type of service or metric, and, possibly, interfaces where measurements are being conducted. Therefore, among other data, the hLS will have a list of IP addresses belonging to the devices that registered with it, and IP addresses of the network devices that are being monitored and which data can be available.

A client computer that aspires to obtain measurement data will query the LS in order to know where this information is available, either in an MA for an already stored measurement, or an MP to conduct a new measurement. If it does not yet know which MA or MP is responsible for the data, the client can contact the gLS, which will respond with a list of hLSs that know about the MA/MPs in question. The client then contacts the hLSs and obtains the URI of the MA or MP service.

For this distributed scheme to operate accordingly, the gLSs must periodically synchronize with each other, and each hLS must publish the data it controls to the upper level gLS. To understand how this information is exchanged, let us consider some cases.

If one gLS contained the complete information published by an hLS, the very purpose of the hLS would be diminished. Instances of gLSs would contain all necessary data, and clients could query in one step. This solution can be said to be, however, less scalable [2], and requires that individual gLSs possess enough computing resources to handle all registering data from all domains monitored by the deployed architecture. A single directory, which is clearly feasible only for restricted domain sizes.

If the gLS were excluded from the picture, clients would not have an "abridged" directory to query. Thus, they would need to contact every hLS in order to find the desired information. An obvious problem in this case is how to let a client know the list of available hLSs. One of the solutions is utilizing upper level summary directories: the gLSs. Other solution would be to have the list of hLSs (such list might become very large) hosted in a web server; the list solves the problem of finding the hLSs, but each of them would still have to be queried until the desired information was found.

In the distributed solution, the gLS therefore contains a summary of the data handled by individual hLSs. Moreover, the local hLS instances must contain enough information to facilitate the discovery of what service data is controlled by them, but should manage the volume of that information that is published into the network; simply publishing the complete stored registrations is inefficient. To achieve this, in the distributed Information Services plane in perfSONAR, the hLS performs a summary of its data and then publishes it to the upper level gLS.

The complete distributed directory algorithm can be realized as having the goal of maintaining sub-directories scattered among domains, and having an abridged directory in the upper level. The task of summarizing the complete base of information handled by all sub-directories is then distributed among instances (the hLSs), and finally the summarized pieces are brought together at the higher level (the gLSs). This method permits higher scalability [2] of the directory service. In particular, we focus on the summarization applied to the hosts' IPv4 addresses.

*3) Summarizing IP Addresses*

IPv4 addresses (henceforth simply "IP" for compactness) are constituted by a sequence of binary numbers. In this vision, we could employ summarizations dealing with IP addresses "beginning" with a specific sequence of bits. This is essentially the CIDR (Classless InterDomain Routing) mechanism (used in IP routing) notation of IP addresses.

In CIDR notation, a portion of the beginning (from left to right) of the IP address is used as an identifier of the network (the network *prefix*), and the rest of the IP address is used to identify a particular host in that network (the *suffix*). A number, written after the IP address and separated by a slash, distinguishes the prefix portion: this number is the number of bits, from the total IP address, reserved for the prefix.

Consequently, we are able to exploit CIDR notation and summarize individual IP addresses into subnets that comprise the appropriate range of addresses. We will see that advertising an IP subnet as summary will not necessarily mean having possession of the *entire* set of hosts in that subnet, but claiming to have *some* hosts in that subnet.

### III. HEURISTIC FOR SUMMARIZING IP ADDRESSES

IP summarization is commonly employed in route advertisement [5], in order to lower the resources needed by routers. In this case, the summarization is generally closely managed and configured by the network manager; in a dynamic information services environment, new services (and their IP addresses) may register frequently. The summarization process must then be automatic. We now analyze aspects that must be taken into account in this procedure, and how we addressed them by means of a heuristic.

#### A. Summarization Goals: the Problem of Balancing Compression and Miss Rate

In the Information Services plane described earlier, IP summarization must fulfill two goals: it must decrease the original set of IP addresses by a reasonable amount (i.e., it must achieve a good compression rate), but it must not summarize so much as to result in claiming many more IP addresses than the original set. If an hLS advertises a summarizing IP of /20, it is claiming to have in its directory all $2^{12}$ hosts in the advertised /20 subnet, even if the hLS holds actually only a small subset of this range. Therefore, claiming a large subnet for a comparable small number of hosts within that subnet poses an extra burden in the search process, because a client will believe that the advertiser hLS does possess all hosts in that subnet (even if it is not the case), and must query the hLS to confirm.

If the desired IP address is not in the hLS, we have a penalty in the form of wasted time and resources to perform the query. This is analogous to a *cache miss*, and the penalty, to a *miss penalty*. If the hLS advertises a smaller subnet, the *precision* of the advertisement improves, thus lowering the probability of a miss (the *miss rate*) and lowering the *miss penalty*. Of course, publishing the complete, not-summarized list of IP addresses incurs in miss penalty zero (a miss rate or probability of zero), but this does not achieve the goal of saving resources in publishing information to the gLSs.

Conversely, advertising the most general summarization, which would be the subnet 0.0.0.0/0 that comprises all possible IP addresses in the IPv4 address space, achieves a maximum compression, but also incurs in the maximum miss rate. In the IP summarization schema, overlapping ranges between hLS can be common; using the least general, tighter subnet possible in the summarization reduces the chances of overlaps, but it also reduces compression.

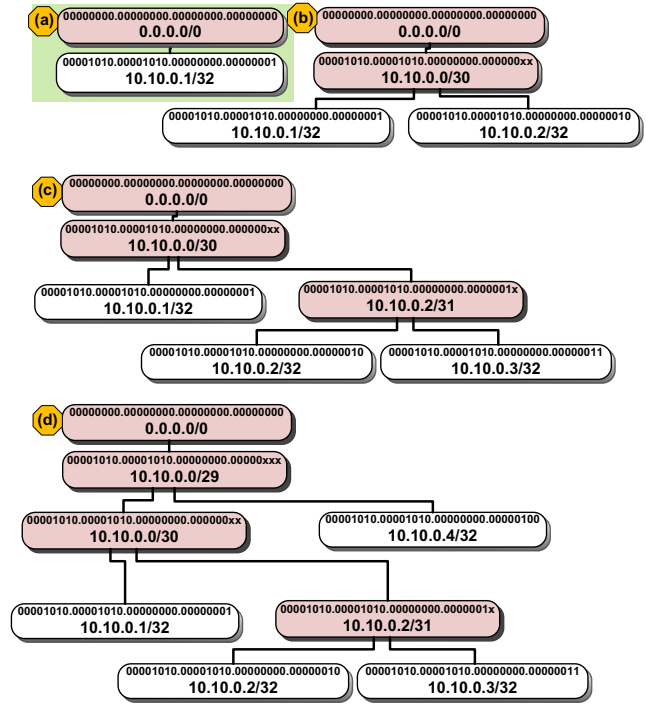

Figure 2. Systematic construction of a PATRICIA tree.

It is clear that IP summarization must then balance compression and miss rate. It cannot pursue maximum compression, as in a lossless file compression system, since that increases the miss probability. Moreover, it cannot focus on minimum miss rate, as that causes low compression rates.

The optimum balance between compression and miss rate is susceptible to administrator interpretation. One might prefer less general IP summarizations and reduce the chance of overlaps, accepting the higher volume of published data and consumed resources. Other might accommodate more general, smaller summarizations and deal with overlaps accordingly. It is unclear, therefore, whether there exists a an optimal IP summarization.

#### B. Finding Subnets: the PATRICIA Data Structure

The key for summarizing a list of IP addresses is finding which subnets the IP addresses match. Then, we pick some of these subnets and some original IP addresses, such that all original IP addresses are represented by some subnet or by themselves, and the final selection fulfills the administrator requirement of compression/miss rate balance.

We attain this process by making use of a special data structure called PATRICIA tree [9][10], which is a form of *trie* [6]. The PATRICIA tree, basically, is a *binary* search tree, where the original data is distributed among the leaves of the tree, and the internal nodes are common string prefixes shared by the respective descendants. Therefore, each internal node of the tree characterizes a whole prefix, and the branches designate different suffixes that exist in the original data set.

The algorithm for generating a PATRICIA tree deems the order of the original data irrelevant. It also guarantees that each node has two children: if, during the construction of the tree, an internal node has only one child, it is simply coalesced into its parent. In our implementation, the internal nodes represent subnet masks in CIDR notation, and the leaves denote the original IP addresses.

Let us follow in Figure 2 the construction of a PATRICIA tree for this given data set: 10.10.0.1, 10.10.0.2, 10.10.0.3, 10.10.0.4. First, all IP addresses are converted into binary form (the binary addresses appear on the top of each decimal address). The tree is created with the common root of all IPv4 addresses: the subnet 0.0.0.0/0. We select the first IP address in the list, 10.10.0.1, and insert it into the tree [Figure 2 (a)] as a child of 0.0.0.0/0 (for the root node alone, the coalescing rule is not followed; it will remain with only one child for now). In this tree, the original IP addresses are represented as /32.

The next IP in the list, 10.10.0.2, shares the same binary prefix with 10.10.0.1 up to the 29$^{th}$ bit. They differ in the last two bits: for 10.10.0.1, they are '01', and for 10.10.0.2, they are '10'. By inserting 10.10.0.2 into the tree [Figure 2 (b)], an ancestor or internal node (or parent) is created for them, containing the common binary prefix for both: 00001010.00001010.00000000.0000000xx, or 10.10.0.0/30.

The address 10.10.0.3 shares the same binary prefix with 10.10.0.2 up to the 31$^{th}$ bit. In Figure 2 (c), we see that a new branch stems from the previous location of 10.10.0.2. It now has the node representing the subnet 10.10.0.2/31 and its two children, 10.10.0.2 and 10.10.0.3.

Finally, in Figure 2 (d), 10.10.0.4 is inserted into the tree. This addresses shares the first 29 bits of prefix with the other /32 addresses already in the tree: a new node (10.10.0.0/29) is created as a parent of 10.10.0.0/30, and 10.10.0.4/32 is added as a child of that new node.

For its essence and characteristics, the PATRICIA data structure successfully favors finding the proper, minimal summarizing nodes (in CIDR notation) for a given set of IP addresses. It remains to elect, from the final tree, the nodes what will constitute the summarized result.

### C. Selecting Nodes to Summarize

In Figure 2 (d), we have a number of options for nodes to pick for summarization. Selecting the root, 0.0.0.0/0, although being a valid summarization, is clearly unwise. Traversing down the tree, we find a next option, the subnet 10.10.0.0/29, that does comprise all original IP addresses. Essentially, after choosing a node as a summarizing spot, we further prune the PATRICIA tree down from that point, since all IP addresses that lie below are already contained within the selected summarizing node.

Other options would be to pick the actual host at 10.10.0.4/32 as a summarizing node, and 10.10.0.0/30 to represent the other three addresses. Alternatively, opt for 10.10.0.4/32 and 10.10.0.1/32, and 10.10.0.2/31 to summarize the remaining two addresses.

*1) The Metrics for Decision: Distance, Density, Minimum Subnet Mask*

The proposed heuristic must decide, without user intervention, which nodes to pick as summarizing nodes from the final conceived PATRICIA tree. We elected three distinct metrics to utilize in the decision procedure: *Mininum Subnet Mask*, *Distance*, and *Density*.

*a) Minimum Subnet Mask*

The *Minimum Subnet Mask* indicates the minimum (not inclusive) acceptable number of bits in the subnet mask (the number after the /), below which a node *will not* be selected as a summarizing node. It is, therefore, a minimum threshold, configurable by the administrator, such that nodes that represent very large subnets (in the view of the administrator) can be avoided as summarizing points. The decision over this metric is detached and takes precedence over the other two metrics. If a node displays a subnet mask of, e.g., size 7, and the Minimum Subnet Mask is set to 8, this node will never be selected as summarizing, regardless of the result of the other two metrics.

*b) Distance*

The *Distance* is the difference, in bits, between a child node's mask (the metric is calculated for each child of a node independently) and its parent's mask. Namely:

$$Distance = ChildMask - ParentMask. \quad (1)$$

This metric expresses the notion of *how many hosts*, or IP addresses, are being *claimed* by a node (if it is made a summarizing node), but *do not actually exist* in the original set. In other words, it conveys the number of CIDR subnets between the current node and the respective child. Since between the parent and the child there is no original IP address, if the parent is made a summarizing node, all the potential IP addresses that can occur between the parent and the child will generate a miss in a search. It is, therefore, beneficial to keep this metric reasonably small.

Observe that the *Distance* does not indicate the *exact* number of IP addresses that are being claimed by a node (if made a summarizing node) and do not actually exist: it is a *notion*. Below a child in the PATRICIA tree, the subnets need not be necessarily complete or filled with all possible addresses.

We use Figure 2 (d) again for an example of *Distance* metric. Take the parent 10.10.0.0/29 and its left child 10.10.0.0/30. The *Distance* is 30 − 29 = 1 bit. This is effectively the minimum possible number between a parent and child, indicating that there are no vacant IP addresses between them. Truly, the 1-bit difference amounts to the following number of addresses between the parent and child:

$$\frac{2^{distance}}{2} = 2^{distance-1} = 2^0 = 1. \quad (2)$$

This is the child address itself. The division by 2 (or the subtraction of 1 from the exponent) in the previous equation is explained by the fact the PATRICIA tree is a binary tree. Each branch of descendants can have half of the total number of descendants.

Between 10.10.0.0/29 and its right child 10.10.0.4/32, we calculate *Distance* = 32 – 29 = 3 bits. We have:

$$\frac{2^{distance}}{2} = 2^{distance-1} = 2^2 = 4. \quad (3)$$

There are, thus, 4 possible addresses in the branch between 10.10.0.0/29 and 10.10.0.4/32. Subtracting the /32 child, we have 3 vacant IP addresses. The value for the *Distance* metric is configurable in the proposed algorithm.

  *c) Density*

The *Density* is a measure of the number of leaves of the current node, divided by the maximum possible number of hosts (IP addresses) below the current node.

$$Density = \frac{\#leaves\ below\ curr.node}{\#max\ leaves\ below\ curr.node}. \quad (4)$$

It is a number between 0 and 1, where 1 is the maximum density, or an indication that all possible IP addresses of a subnet are taken. To calculate this metric, the algorithm keeps track of the number of leaves below a node (i.e., the number of hosts below a node). The maximum number of hosts, or leaves, or original IP addresses, which can exist below the current node, can be computed by subtracting the current node's mask (bits) from the maximum mask (bits), and using this as the exponent of 2. Two to the power of this exponent will yield the maximum number of leaves below the node. The maximum mask depends on the address system at use: for IPv4, the maximum mask is 32 bits. Hence, if *MaxMask* is the maximum mask in bits, and *CurrentNodeMask* is the current node's mask in bits, we have:

$$Density = \frac{\#leaves\ below\ curr.node}{2^{MaxMask-CurrentNodeMask}}. \quad (5)$$

As the PATRICIA tree is constructed, the algorithm updates the *Density* metric for each node and stores this information within the node's data structure. Later, this data will be retrieved in order to decide whether a node is to be made a summarizing node, or not. We now investigate how the heuristic operates this decision.

  *2) Using the Metrics to Decide Summarizing Nodes*

The algorithm that implements the heuristic begins by traversing the PATRICIA tree in order (visit root, navigate left branch, navigate right branch), recursively. If the node is a leaf, then make it a summarizing node (since there is nothing below it) and return. Otherwise, verify whether the current node's mask is smaller than the *Minimum Subnet Mask*. If it is, continue traversing the tree in order (i.e., the current node denotes a subnet that is too large).

Else, calculate the *Distance* of the current node from each of its children. If any *Distance* is greater than an administrator-defined value, then continue traversing the tree in order (i.e., summarizing at the current node would represent too many vacant IP addresses).

Else, retrieve the *Density* of each of the current node's children. If either is smaller than an administrator-defined threshold, then continue traversing the tree in order (i.e., there are few leaves in the left or right branch, so we attempt to summarize closer to the leaves in this case). Notice that the heuristic checks the child's *Density* (and for each child independently), and not the current node's *Density*. This procedure conveys a better notion for the case of highly unbalanced trees, or trees where one branch is dense, whereas the other branch is comparatively sparse.

Else, make the current node a summarizing node; prune the tree at this spot, and return.

*Density* and *Distance* are somewhat related metrics; both essentially express the size of the set of IP addresses that could be summarized, and how many vacant addresses would be summarized altogether. One, however, is more biased towards the vacant addresses than the other is, and we have found in our experiments that we can have a stronger control of the summarization by combining the decision upon both metrics.

  *3) Administrator-Defined Configuration: The Granularity and Minimum Subnet Mask*

In the procedure described, the administrator can configure three values that the heuristic employs to decide the summarizing nodes. By tuning these values, the administrator can achieve a coarser summarization (one with a smaller number of results, or more compression), or a finer one (with a larger number of results, or less compression).

To simplify this configuration, we have conceived a single setting, called *granularity*, which is capable of regulating both *Distance* and *Density* to their appropriate individual settings. The *granularity* assumes an integer value from 0 to 3, where 0 corresponds to the finer, or less compressed summarization, and 3 implies the coarser, more compressed summarization. We keep the *Minimum Subnet Mask* setting apart, because this threshold is usually less frequently changed, even among different compression choices. Administrators can profit in setting this separately.

TABLE I. MAPPING OF THRESHOLD VALUES USED BY GRANULARITY.

| Granularity | Distance | Density |
|---|---|---|
| 0 | 4 | 1e-5 |
| 1 | 8 | 1e-6 |
| 2 | 12 | 1e-7 |
| 3 | 16 | 1e-8 |

The *granularity* works by mapping, or converting, its value to the specific value of *Density* and *Distance* (in bits). We therefore have selected, from each metric, four fixed values that are able to yield the four different compressions. Each of those numbers was reached by experimentation on a test bed, and by our interpretation of reasonable summarizing results. Table I brings the final mapping.

Figure 3 shows the graph *Distance x Granularity*. It is clearly a simple linear relation. Figure 4 depicts the relationship *Density x Granularity*, which resembles an inverse exponential curve. Using regression analysis, we can appropriately assign an interpolating equation to each metric. For *Distance*:

$$y = 4x + 4. \quad (6)$$

And, for *Density*:

$$y = 10^{-5}e^{-2.303x}. \qquad (7)$$

Where *x* is the *Granularity*, and *y* is the metric.

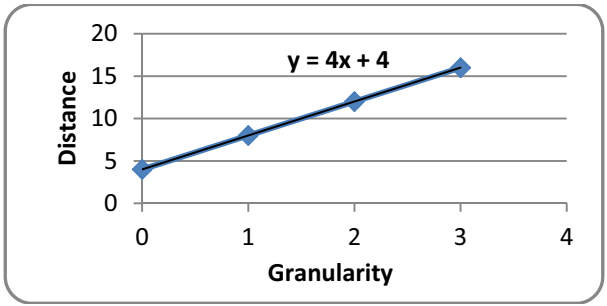

Figure 3. Graph showing the relationship *Distance x Granularity*.

The algorithm applies these equations to convert from current retrieved *densities* and calculated *distances* to a *granularity* value. Then, it tests the thresholds as explained in subsection (III.C.2), and decides whether to choose a summarizing node.

The last step of the algorithm is performed by the higher-level instances, which is combining the individual summarized pieces into one abridged list.

## IV. EXPERIMENTAL RESULTS

We investigated the proposed heuristic and implementation employing the simulated distributed information services test bed portrayed in Figure 5. In this test bed, there are nine hLSs; each of them receives registrations from services and stores their IP addresses. Then, they individually run the algorithm to summarize these addresses and publish their results to the gLS, which in turn merge the parts into one list.

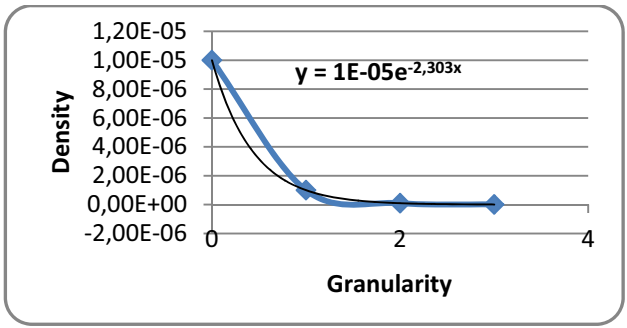

Figure 4. Graph showing the relationship *Density x Granularity*.

Each hLS manages IP addresses extracted from real networks [12] that utilize the Information Services plane in the perfSONAR system (there is one duplicate IP address, located in APAN and ESnet hLSs). The goal is to derive the performance of the heuristic, in particular how much of each original subset is compressed by the summarization technique and the final merged compression. We also verify the performance when only one hLS contained all subsets, and then performed one overall summarization. Table II compiles the results from the nine individual hLSs and the final merged list of addresses. For each hLS (which are named according to the real network domain they simulate), the table displays the original list size each hLS processed in number of IP addresses, the final summarized size (also in number of IP addresses) and compression rate (summarized size divided by original size, or simply the percentage of the original size that the summarized set represents) for distinct *granularity* settings. The *Minimum Subnet Mask* was always set to 8 (signifying that nodes with a mask from 8 and below were not accepted as summarizing nodes) for these tests, based on our preference that no summarized IP address comprised subnets larger than the /8 mask.

The last row of Table II indicates the previous statistics calculated for the "big" list in the gLS, built by merging the published summarized sub-lists of each hLS.

### A. Analysis

The results demonstrate that the summarization performance is dependent on the input data set. Some hLSs, for example, summarized to the same compression rate regardless of the selected *granularity* mode. Data sets might exhibit a particular configuration or allotment of IP addresses that hurts the flexibility of the choice of summarizing nodes. This feature of some data sets may account for the behavior displayed by some hLSs in the test bed, which summarized similarly for different granularities.

The hLS responsible for the larger data set, ESnet, could summarize it to less than 10% of its original size. The *granularity* setting is capable of changing the compression rate accordingly; the amount of this change, however, is dependent on the conformation of the data set, as explained earlier.

Overall, the compression rate on the experimental test bed ranged below 30% (the lower the number, higher the compression is). Roughly, a compression rate of 30% on the list of IP addresses, for our distributed IP summarization algorithm, means that a higher-level directory may contain only 30% of the original IP address list size, and still carry comparable (within certain precision) information for the purposes of the distributed, hierarchical directory in the Information Services.

The last row of Table II shows the results from the point of view of the higher-level gLS. Again, this gLS did not summarize; it received the already summarized pieces from the hLSs and merged them. For this gLS, the overall compression rate ranged from 13% for granularity 0, to 6.9% for granularity 3.

The final experiment consisted of adjusting the test bed such that a single hLS received all registrations from all services. Thus, this hLS contained all the IP addresses from all domains combined. It summarized this unique, combined list and published to the gLS. Table III tabulates the results; the *distributed* row copies the last row of Table III having the final summarization results from the algorithm running in the distributed environment, as viewed from the gLS. The *single* row illustrates the outcome when a single instance of the hLS now summarizes all the existing IP addresses as one combined input. The *decrease* row informs the reduction in the summarized set size, per granularity, from the *single* to

[Should be Table II.]

the *distributed* operation (1 – [*single* final size/*distributed* final size]).

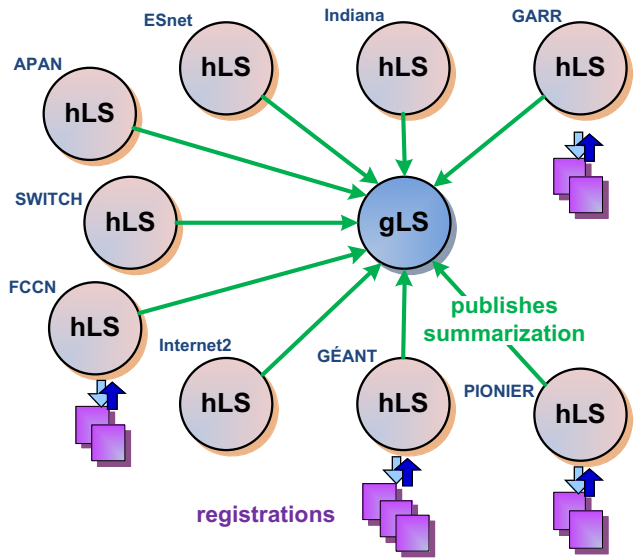

Figure 5. Simulated distributed test bed for validating the summarization algorithm.

According to Table III, in the *single* operation the gLS now sees a more compressed list of addresses as compared to the distributed test bed, given the same input of IP addresses. Although smaller, this new list loses precision in contrast with the one generated by the distributed instances of the algorithm running in the hLSs. The heuristic is attempting to summarize a large set made of subsets, where each subset has a particular pattern of IP addresses (i.e., enclose particular subnets), and is then selecting summarizing IP addresses that represent larger portions of these subnets. Obviously, not pondering scalability issues, if the single hLS could divide the combined set into the same original subsets from the distributed test bed, and employed the heuristic individually in each subnet, the result would be the same as the distributed operation.

## V. OTHER SUMMARIZATION TECHNIQUES AND ROUTE AGGREGATION ALGORITHMS

Currently, the LS incorporates an algorithm to perform IP summarization that relies on a voting scheme in order to identify the subnets that represent most of the original IP addresses [12]. For each original address, that algorithm expands all subnets comprising the address, storing them into a list. If a subnet was already expanded by a previous address, its vote counter is incremented. The algorithm then selects candidates for summarizing addresses by picking subnets that have at least one original, /32 IP address child. The user cannot influence this decision, and the final summarizing subnets might be of any size. Distinctively, our heuristic allows for control of the compression level of the summarization, and also implements mechanisms to avoid selecting summarizing subnets that might be deemed very large.

We have also surveyed a number of route aggregation algorithms, notably [4][5][11][13][15]. While they are related, a direct comparison between our heuristic and those references is not possible. The main objective of those efforts is IP lookup performance improvement for routing, and they utilize the "next hop" information in decision making. Our algorithm is not intended to be used for routing, and in fact the "next hop" information is not available as an input for the purposes of the summarization.

## VI. CONCLUSION

We have demonstrated, in this paper, a heuristic applied in a distributed Information Services architecture to construct a summarized set from original sets of IPv4 addresses. These addresses can be typically services registering to a hierarchical, distributed directory service.

As motivators, we outlined the needs of the Information Services plane adopted in perfSONAR, a performance monitoring architecture. This Information Services plane includes a distributed directory service, where other services register to and clients perform queries to find other services and performance data. To operate adequately, the service relies on lower level instances, which publish a summary of their controlled information to the upper level instance layer. By employing summarization, the service administers the volume of information that is published into the network. Moreover, resources, such as memory and storage, can be constrained.

We further described the mechanics of IP summarization and the techniques utilized by our proposed heuristic to obtain a final summarization. Essentially, our heuristic assembles a special data structure, called a PATRICIA tree; then, it selects nodes from this tree based on specific metrics and configured thresholds. Namely, we use *distance, density*, and *minimum subnet mask*. This selection follows particular interpretations of appropriate summarizations. Then, distributed sets of summarized addresses are assembled into one abridged set.

The heuristic was applied in a simulated distributed information services test bed, comprised of instances of directory services. The instances received registrations of IP addresses collected from real networks that adopt perfSONAR architecture, and published their outputs to a higher instance that merged them. The results illustrate compressions rates below 30% (the smaller, the more compressed) and the capability of adjusting the compression rate by means of a *granularity* setting. The advantage of a compressed set is that less resources are required to store and transmit it, yet the set still conveys comparable information (within adjustable bounds) to the original set. It was also verified that the heuristic performance is evidently dependent on the arrangement of IP addresses in the original data set.

TABLE II. RESULTS OF SIMULATED DISTRIBUTED INFORMATION SERVICES RUNNING THE IP SUMMARIZATION HEURISTIC.

| set | Original Set Size (IP addresses) | Summarized Size and Compression Rate | | | | | | | |
|---|---|---|---|---|---|---|---|---|---|
| | | Granularity 0 | | Granularity 1 | | Granularity 2 | | Granularity 3 | |
| | | Final Size | Comp. Rate | Final Size | Comp. Rate | Final Size | Comp. Rate | Final Size | Comp. Rate |
| APAN | 104 | 19 | 0,18269231 | 17 | 0,16346154 | 17 | 0,16346154 | 13 | 0,125 |
| ESnet | 618 | 58 | 0,09385113 | 28 | 0,04530744 | 20 | 0,03236246 | 20 | 0,03236246 |
| FCCN | 43 | 4 | 0,09302326 | 4 | 0,09302326 | 4 | 0,09302326 | 3 | 0,06976744 |
| GARR | 163 | 2 | 0,01226994 | 2 | 0,01226994 | 2 | 0,01226994 | 2 | 0,01226994 |
| GÉANT | 30 | 5 | 0,16666667 | 5 | 0,16666667 | 5 | 0,16666667 | 5 | 0,16666667 |
| Internet2 | 282 | 86 | 0,30496454 | 54 | 0,19148936 | 44 | 0,15602837 | 42 | 0,14893617 |
| Indiana | 79 | 14 | 0,17721519 | 12 | 0,15189873 | 12 | 0,15189873 | 10 | 0,12658228 |
| PIONIER | 27 | 5 | 0,18518519 | 5 | 0,18518519 | 5 | 0,18518519 | 3 | 0,11111111 |
| SWITCH | 214 | 15 | 0,07009346 | 12 | 0,05607477 | 10 | 0,04672897 | 10 | 0,04672897 |
| Final | 1560 | 208 | 0,13333333 | 139 | 0,08910256 | 119 | 0,07628205 | 108 | 0,06923077 |

*Comp. Rate = Compression Rate = summarized / original (lower is more compressed); **Minimum Subnet Mask = 8

TABLE III. RESULTS OF SIMULATED DISTRIBUTED INFORMATION SERVICES RUNNING THE IP SUMMARIZATION HEURISTIC IN DISTRIBUTED MODE, AND SINGLE MODE.

| hLS | Original Set Size (IP addresses) | Summarized Size and Compression Rate | | | | | | | |
|---|---|---|---|---|---|---|---|---|---|
| | | Granularity 0 | | Granularity 1 | | Granularity 2 | | Granularity 3 | |
| | | Final Size | Comp. Rate | Final Size | Comp. Rate | Final Size | Comp. Rate | Final Size | Comp. Rate |
| Distributed | 1560 | 208 | 0,1333333 | 139 | 0,0891026 | 119 | 0,0762821 | 108 | 0,0692308 |
| Single | 1560 | 180 | 0,1153846 | 102 | 0,0653846 | 83 | 0,0532051 | 76 | 0,0487179 |
| Decrease | | 13,46% | | 26,62% | | 30,25% | | 29,63% | |

*Comp. Rate = Compression Rate = summarized / original (lower is more compressed); **Minimum Subnet Mask = 8
***Decrease = 1 - (single/distributed final size)